\documentclass[showpacs,showkeys,preprintnumbers, amsmath,amssymb,preprint]{revtex4-1}
\usepackage{graphicx}
\usepackage{epsfig}
\usepackage{bm}
\usepackage{arcs}
\usepackage{fontenc}
\usepackage{graphicx}
\usepackage[dvips]{hyperref}
\usepackage{bm}
\usepackage{amsmath}
\usepackage{tipa}
\usepackage{MnSymbol}
\usepackage{color}
\usepackage{soul}

\begin{document}

\title{Teleparallel Equivalent of Lovelock Gravity }
\author{P. A. Gonz\'{a}lez}
\email{pablo.gonzalez@udp.cl}
\affiliation{Facultad de Ingenier\'{\i}a, Universidad Diego Portales, \\
Avenida Ej\'{e}rcito Libertador 441, Casilla 298-V, Santiago, Chile.}
\author{Yerko V\'{a}squez.}
\email{yvasquez@userena.cl}
\affiliation{Departamento de F\'{\i}sica, Facultad de Ciencias, Universidad de La Serena,\\
Avenida Cisternas 1200, La Serena, Chile.}
\date{\today }

\begin{abstract}

There is a growing interest in modified gravity theories based on torsion, as these theories exhibit interesting cosmological implications. In this work, inspired by the teleparallel formulation of general relativity, we present its extension to Lovelock gravity known as the most natural extension of general relativity in higher-dimensional space-times. First, we review 
the teleparallel equivalent of general relativity and Gauss-Bonnet gravity, and then we construct 
the teleparallel equivalent of Lovelock gravity. In order to achieve this goal we use the vielbein and the connection without imposing the Weitzenb{\"o}ck connection. Then, we extract the teleparallel formulation of the theory by setting the curvature to null.

\end{abstract}
\maketitle
\section{Introduction}

Higher-order curvature theories of gravity lead to a wide variety of alternative theories of gravity with a rich phenomenology. Among them, the Lovelock theory is the most natural extension of general relativity in higher-dimensional space-times that generates second-order field equations  \cite{Lovelock:1971yv}. Remarkably, the action contains terms that appear as corrections to the Einstein-Hilbert action in the context of string theory. Supersymmetric extension, exact black hole solutions, scalar perturbations, thermodynamic, holographic aspects and other properties of Lovelock gravity have been extensively studied over time.

On the other hand, extensions and generalizations of theories based on torsion have gained a lot of attention too.  In particular, the so-called ``teleparallel equivalent of general relativity'' (TEGR) \cite{Unzicker:2005in, Hayashi:1979qx} is an
equivalent formulation of gravity; however, instead of using the curvature
defined via the Levi-Civita connection, it uses the Weitzenb{\"o}ck connection that has no curvature, only torsion \cite{Aldrovandi:2013wha}.  A natural extension of the TEGR is the so-called $f(T)$ gravity, which is represented by a function of the torsion scalar $T$ as Lagrangian density \cite{Ferraro:2006jd, Ferraro:2008ey, Bengochea:2008gz,Linder:2010py}.
The $f(T)$ theories pick up preferred referential frames which constitute the autoparallel curves of the given manifold. 
A genuine advantage of $f(T)$ gravity  compared with other deformed gravitational schemes is that the differential equations for the vielbein components are second-order differential equations. However, the effects of the additional degrees of freedom that certainly exist in $f(T)$ theories is a consequence of breaking the local Lorentz invariance that  these theories exhibit. Nevertheless, it was found that on the flat FRW background with a scalar field,  up to second-order linear perturbations do not reveal any extra degree of freedom at all \cite{Izumi:2012qj}. As such, it is fair to say  that the nature of these additional degrees of freedom remains unknown. Remarkably, it is possible to modify $f(T)$ theory in order to make it manifestly a Lorentz invariant. However, it will generically have different dynamics and will reduce to $f(T)$ gravity in some local Lorentz frames \cite{Li:2010cg, Weinberg, Arcos:2010gi}.
Clearly, by extending this
geometry sector, one of the goals is to solve the puzzle of dark energy and
dark matter without asking for new material ingredients that have
not yet been detected  by  experiments \cite{Capozziello:2007ec,Ghosh:2012pg}. For instance, a Born-Infeld $f(T)$ gravity Lagrangian was used to address the physically inadmissible divergencies occurring in the standard cosmological Big Bang model, rendering the space-time geodesically complete and powering an inflationary stage without the introduction of an inflaton field \cite{Ferraro:2008ey}. Also, it is believed that $f(T)$ gravity could be a reliable approach to address the shortcomings of general relativity at high-energy scales \cite{Capozziello:2011et}.  Furthermore, both inflation and the dark energy-dominated stage can be realized in the Kaluza-Klein and Randall-Sundrum models, respectively \cite{Bamba:2013fta}.
In this way, $f(T)$ gravity has gained attention and
has been proven to exhibit interesting cosmological implications. On the other hand, the search for black hole solutions in $f(T)$ gravity is not a trivial problem, and there are only a few exact solutions, see for instance \cite{G1, solutions,Rodrigues:2013ifa}.

Furthermore, generalizations such as the teleparallel equivalent of Gauss-Bonnet gravity \cite{Kofinas:2014owa, Kofinas:2014daa}, the Kaluza-Klein theory for teleparallel gravity \cite{Geng:2014nfa} and scalar-torsion gravity theories \cite{Geng:2011aj, Kofinas:2015hla, Kofinas:2015zaa} have been of recent interest as these theories have been proven to exhibit interesting cosmological implications. For instance,  
modified gravity theories based on $f(T, T_G)$, where $T_G$ is the
torsion invariant and is equivalent to the Gauss-Bonnet term, provides a unified description of the cosmological history from early-times inflation to late-times self-acceleration without the inclusion of a cosmological constant for some cases of $f(T, T_G)$ \cite{Kofinas:2014daa}.
Besides, it was shown that in three-dimensional teleparallel gravity, there are asymptotically AdS black hole solutions with a scalar field non-minimally coupled to gravity with a self-interacting potential \cite{Gonzalez:2014pwa}, where the diagonal frame used parallelizes the space-time, that is, the frame defines a global set of bases covering the whole tangent bundle \cite{Fiorini:2013hva}. The main purpose of this work is to present the teleparallel equivalent of Lovelock gravity inspired by Lovelock gravity known as the most natural extension of general relativity in higher dimensional space-times and by the recent generalizations of TEGR.

The paper is organized as follows. In Sec. II we give a brief review of
Lovelock gravity. Then, in Sec. III we give a brief review of the teleparallel equivalent of general relativity.
In Sec. IV, we construct the teleparallel equivalent of general relativity. Then, we construct the teleparallel equivalent of Gauss-Bonnet gravity, and finally we construct the teleparallel equivalent of Lovelock gravity. Then, we conclude in Sec. V.

\section{Lovelock Gravity}

The Lanczos-Lovelock action is the most natural extension of general relativity in higher-dimensional space-times that generate second-order field equations. This action is non-linear in the Riemann tensor, and it differs from the Einstein-Hilbert (EH) action only if the space-time has more than four dimensions. In $D-$dimensions it can be written as follows
\begin{equation}\label{dtopologicalaction}\
I_{k}=\kappa \int \sum_{q=0}^{k}c_{q}^{k}L_{(q)}~,
\end{equation}%
with
\begin{equation}\label{polinomio}\
L_{(q)}=\epsilon_{a_1...a_D}R^{a_1 a_2}\wedge...\wedge R^{a_{2q-1} a_{2q}}\wedge e^{a_{2q+1}}\wedge...\wedge e^{a_{D}}~,
\end{equation}%
where $1\leq k\leq [\frac{D-1}{2}]$ ($[x]$ denotes the integer part of $x$), $c_{q}^{k}=\frac{\ell^{2(q-k)}}{D-2q}(^{k}_{q})$ for $q\leq k$ and vanishes for $q>k$, $e^a$ stands for the vielbein 1-form, $R^{a b}$ stands for the curvature 2-form, and $\kappa$ and $\ell$ are
related to the gravitational constant $G_{k}$ and the cosmological
constant $\Lambda$ through
\begin{equation}\label{definitionk}\
\kappa=\frac{1}{2(D-2)!\Omega _{D-2}G_{k}}~,
\end{equation}%
\begin{equation}\label{lambda}\
\Lambda=-\frac{(D-1)(D-2)}{2\ell^2}~,
\end{equation}%
where $\Omega _{D-2}$ corresponds
to the volume of a unit $(D-2)$-dimensional sphere. The following field equations are obtained when varying with respect to the vielbein $e^{a}$ and the connection $\omega^{ab}$, respectively
\begin{equation}
\epsilon_{aa_1 \cdot\cdot\cdot a_{D-1}}\tilde{R}^{a_1a_2}\wedge \cdot\cdot\cdot \wedge \tilde{R}^{a_{2k-1}a_{2k}}\wedge e^{a_{2k+1}}\wedge \cdot\cdot\cdot \wedge e^{a_{D-1}}=0~,
\end{equation}
\begin{equation}
\epsilon_{aba_3\cdot\cdot\cdot a_{D}}\tilde{R}^{a_3a_4}\wedge \cdot\cdot\cdot \wedge \tilde{R}^{a_{2k-1}a_{2k}}\wedge T^{a_{2k+1}}\wedge e^{a_{2k+2}}\wedge \cdot\cdot\cdot \wedge e^{a_{D-1}}=0~,
\end{equation}
where $\tilde{R}^{ab}:=R^{ab}+\frac{1}{\ell^2}e^a \wedge e^b$ and $T^a$ is the torsion 2-form. The theory with $k=2$ is described by a Lagrangian which is a linear combination of Gauss-Bonnet density, the EH Lagrangian and the volume term with fixed weights \cite{Crisostomo:2000bb}. Static black
hole-like geometries with spherical topology were
found~\cite{Crisostomo:2000bb} to possess topologically non-trivial
AdS asymptotics.  These theories and their corresponding solutions
were classified by an integer $k$, which corresponds to the
highest power of curvature in the Lagrangian. If $D-2k=1$, the
solutions are known as Chern-Simons black holes (for a review on
the Chern-Simons theories see \cite{Zanelli:2005sa}). These
solutions were further generalized to other
topologies~\cite{Aros:2000ij} and can be described in general
by a non-trivial transverse spatial section $\sum_{\gamma}$ of
$(D-2)$-dimensions  labelled by the constant $\gamma=+1, -1, 0$, which represents the curvature of the transverse section,
corresponding to a spherical, hyperbolic or plane section,
respectively. On the other hand, exact black hole solutions with non-maximally symmetric horizons have been found  in the context of third-order Lovelock gravity
\cite{Farhangkhah:2014zka}.
 
\section{Teleparallel Equivalent of General Relativity } 

In 1928 Einstein proposed the idea of teleparallelism to unify gravity and electromagnetism into a unified field theory, which corresponds to an equivalent formulation of general relativity nowadays known as teleparallel equivalent of general relativity \cite{Unzicker:2005in, Hayashi:1979qx}, where the Weitzenb\"{o}ck connection is used  to define the covariant derivative instead of the Levi-Civita connection, which is used to define the covariant derivative in the context of general relativity. The Weitzenb\"{o}ck connection has non-null torsion; however, it is curvatureless, which implies that this formulation of gravity exhibits only torsion. Thus,  in TEGR the torsion tensor include all the information concerning the
gravitational field and the action is given by
\begin{eqnarray}  \label{action0}
S = \frac{1}{16 \pi G}\int d^4x e \left(T+\mathcal{L}_{m}\right)~,
\end{eqnarray}
where $G$ is the Newton constant, $e = \text{det}(e_{\,\, \mu}^a) = \sqrt{-g}$, $T$ is the torsion scalar and $%
\mathcal{L}_{m}$ stands for the matter Lagrangian. Actually, $T$ is the result of a very specific quadratic combination of irreducible representations of the torsion tensor under the Lorentz group $SO(1,3)$ \cite{Hehl:1994ue}. 
The equations of motion can be obtained through the variation of the action (\ref{action0}) with respect to the vielbein, which yields
\begin{eqnarray}\label{eom}
e^{-1}\partial_{\mu}(eS_{a}{}^{\mu\nu})
-e_{a}^{\,\, \lambda}T^{\rho}{}_{\mu\lambda}S_{\rho}{}^{\nu\mu}
-\frac{1}{4}e_{a}^{\,\, \nu
}T
= 4\pi Ge_{a}^{\,\, \rho}\overset {\mathbf{em}}T_{\rho}{}^{\nu}~,
\end{eqnarray}
where the mixed indices are used as in $S_a{}^{\mu\nu} =
e_a^{\,\, \rho}S_{\rho}{}^{\mu\nu}$. The vielbein field ${\mathbf{e}_a(x^\mu)}$ forms an orthonormal
basis for the tangent space at each point $x^\mu$ of the manifold, that is
$\mathbf{e} _a\cdot\mathbf{e}_b=\eta_{ab}$, with
$\eta_{ab}=diag (1,-1,-1,-1)$. Moreover, the vector $\mathbf{e}_a$ can
be expressed in terms of its components $e_a^{\,\, \mu}$ on a coordinate basis,
namely $\mathbf{e}_a=e^{\,\, \mu}_a\partial_\mu$. The torsion scalar is given by $T\ =\ S_{\ \mu \nu }^{\rho }\ T_{\rho }^{\ \mu \nu }\, ,  \label{Weitinvar}$
where $T_{\rho }^{\,\,\,\mu \nu}$ are the components of the torsion 2-form
$T^{a}=de^{a}$ coming from the Weitzenb\"{o}ck connection $\Gamma^\lambda_{\nu%
\mu}=\,e_{a}^{\,\, \lambda }\,\partial _{\nu }e_{\,\, \mu }^{a}$. The dual vielbein $e^{a}(x^{\mu})$ denotes the dual base of ${\mathbf{e}_a(x^\mu)}$ for the cotangent space at each point $x^\mu$ of the manifold, and $S_{\lambda \mu
\rho }$ is defined according to
\begin{equation}
S_{\ \mu \nu }^{\rho }=\frac{1}{4}\,(T_{\ \mu \nu }^{\rho }-T_{\mu \nu }^{\
\ \ \rho }+T_{\nu \mu }^{\ \ \ \rho })+\frac{1}{2}\ \delta _{\mu }^{\rho }\
T_{\sigma \nu }^{\ \ \ \sigma }-\frac{1}{2}\ \delta _{\nu }^{\rho }\
T_{\sigma \mu }^{\ \ \,\sigma }~.  \label{tensor}
\end{equation}
Note that the tensor
$\overset{\mathbf{em}}{T%
}_{\rho}{}^{\nu}$ on the right-hand side is the usual energy-momentum
tensor.
Furthermore, the metric and the dual vielbein are related by
\begin{equation}\label{cnn}
g_{\mu \nu}=e^{a}_{\,\, \mu}e^{b}_{\,\, \nu} \eta^{ab}~,
\end{equation}
and the torsion scalar $T$ and the Levi-Civita Ricci scalar $\bar{R}$ are related by the equation:
\begin{equation}\label{dif}
eT=-e\bar{R}+2\partial _{\nu} \left( eT_{\sigma}^{\,\,\,\ \sigma \nu} \right)~;
\end{equation} 
therefore, the theory is called ``teleparallel equivalent of general relativity'' due to the equations of motion (\ref{eom}) being exactly the same as those of general relativity for every geometry choice because the Lagrangians of both theories differ by just one boundary term as shown in equation (\ref{dif}).
\section{Teleparallel Equivalent of Lovelock Gravity}
\label{section}

In this section, we will construct the teleparallel equivalent of Lovelock gravity using the approach of \cite{Kofinas:2014owa}. To achieve the goal, we will begin by considering the Lanczos-Lovelock action and then we will extract the teleparallel equivalent of Lovelock gravity $T_L$ by setting the curvature to null. 
So, without imposing the Weitzenb{\"o}ck connection from the beginning, it is possible to define the torsion 2-form as
\begin{equation}
T^a=de^a+\omega^a_{\,\ b}\wedge e^b~,
\end{equation}
and the curvature 2-form as
\begin{equation}
R^a_{\,\ b}=d\omega^a_{\,\ b}+\omega^a_{\,\ c}\wedge\omega^c_{\,\ b}~.
\end{equation}
The dynamic variables are the vielbein $e^a$ and the connection $\omega^a_{\,\ b}$ 1-forms. The curvature 2-form corresponding to the torsionless Levi-Civita spin connection $\bar{\omega}^a_{\,\ b}$ is denoted by $\bar{R}^a_{\,\ b}$
\begin{equation}
\bar{R}^a_{\,\ b}=d \bar{\omega}^a_{\,\ b}+\bar{\omega}^a_{\,\ c}\wedge \bar{\omega}^c_{\,\ b}~.
\end{equation}
The arbitrary spin connection $\omega^a_{\,\ b}$ is then related to $\bar{\omega}^a_{\,\ b}$ through the relation
\begin{equation}\label{contortion}
K_{ab}=-K_{ba}=\omega_{ab}-\bar{\omega}_{ab}~,
\end{equation}
where $K_{ab}$ denotes the contortion 1-form. On the other hand, the covariant exterior derivative $D$ of the connection $\omega_{ab}$ acts on a set of $p$-forms $\phi^a _{\,\ b}$ as $D\phi^a_{\,\ b}=d\phi^a_{\,\ b}+\omega^a_{\,\ c}\wedge\phi^c_{\,\ b}-(-1)^p\phi^a_{\,\ c}\wedge\omega^c_{\,\ b}$, whereas the covariant derivative $\bar{D}$ is defined in a similar manner for the Levi-Civita connection $\bar{\omega}_{ab}$, being $\bar{T}^a=\bar{D}e^a=0$. Then, using (\ref{contortion}) it is possible to write the curvature in terms of the Riemannian curvature and the contortion
\begin{equation} \label{curvatura}
R^{ab}=\bar{R}^{ab}+\bar{D}K^{ab}+K^a_{\,\ c}\wedge K^{cb}~.
\end{equation}
Also, we have the Bianchi identities: $DT^a=R^a_{\,\ b}\wedge e^b$ and $DR^a_{\,\ b}=0$. Another useful relation we will employ on the next sections is $D^2\phi^a_{\,\ b}=R^a_{\,\ c}\wedge\phi^c_{\,\ b}-\phi^a_{\,\ c}\wedge R^c_{\,\ b}$ \cite{Kofinas:2014owa}. Also, the covariant derivative of the contortion is given explicitly by $DK^a_{\,\ b}=dK^a_{\,\ b}+\omega^a_{\,\ c}\wedge K^c_{\,\ b}+K^a_{\,\ c}\wedge\omega^c_{\,\ b}$. Furthermore, the following useful expressions will be used in the next sections:
\begin{eqnarray} \label{expressions}
\notag \bar{D}K^a_{\,\ b} &=& DK^a_{\,\ b}-2K^a_{\,\ c}\wedge K^c_{\,\ b}~, \\
\bar{R}^{ab} &=& R^{ab}+K^a_{\,\ c}\wedge K^c_{\,\ b}-DK^a_{\,\ b}~.
\end{eqnarray}
In order to simplify the writing of equations we will use the following notation:
\begin{eqnarray} \label{notation}
\notag R^n \wedge \langle KK \rangle ^m \wedge E ^{D-2n-2m}&=&\epsilon_{a_1 a_2 \cdot\cdot\cdot  a_D}R^{a_1 a_2}\wedge\cdot\cdot\cdot \, \wedge R^{a_{2n-1}a_{2n}}\wedge K^{a_{2n+1} c_1}\wedge K^{\, \, \, a_{2n+2}}_{ c_1}\wedge \cdot\cdot\cdot \\
&&\wedge K^{a_{2n+2m-1} c_{2m-1}}\wedge K^{\, \, \, \, \, \, a_{2n+2m}}_{ c_{2m-1}}\wedge e^{a_{2n+2m+1}}\wedge\cdot\cdot\cdot\, \wedge e^{a_D}~,
\end{eqnarray}
where $R^n$ denotes the first $n$ products of curvature 2-forms $R^{a_1 a_2}\wedge\cdot\cdot\cdot \, \wedge R^{a_{2n-1}a_{2n}}$, $\langle KK \rangle ^m$ denotes the next $m$ products of terms of the form $K^{a_i c}\wedge K_c^{\,\ a_{i+1}}$, where the brackets $ \langle \rangle$ mean that one index is contracted, and the term $E^{D-2n-2m}$ denotes the remaining products of the $D-2n-2m$ vielbeins $e^{a_{2n+2m+1}}\wedge\cdot\cdot\cdot\, \wedge e^{a_D}$. Therefore, using this notation, Equation (\ref{curvatura}) reads $R=\bar{R}+\bar{D}K+\langle KK \rangle$; it must be noted that expressions like this only make sense when they are contracted with the Levi-Civita symbol according to the definition (\ref{notation}), and 
it is possible to write (\ref{polinomio}) compactly as
\begin{eqnarray}\label{polimonioT}
\notag \mathcal{L}_{(n)}&=&R^{n}\wedge E^{D-2n} \\ 
&=&(\bar{D}K+\bar{R}+\langle KK \rangle )^{n}\wedge E^{D-2n}~.
\end{eqnarray}
In the next sections, we will consider the above notation and we will study in some detail the case $n=1$ that corresponds to the teleparallel equivalent of general relativity. Then, we will study the case $n= 2$ that corresponds to the teleparallel equivalent of Gauss-Bonnet gravity, and to end, to generalize the study we will consider the teleparallel equivalent of Lovelock gravity that corresponds to an arbitrary $n$.

\begin{itemize}

\item {Teleparallel equivalent of general relativity}

For $n=1$, the Lagrangian (\ref{polimonioT}) corresponds to the Einstein-Cartan Lagrangian $ \mathcal{L}_{(1)} $, and we have:
\begin{eqnarray} \label{Lagrangian1}
\notag \mathcal{L}_{(1)}&=& R \wedge E^{D-2} \\ 
&=&(\bar{D}K+\bar{R}+\langle KK \rangle )\wedge E^{D-2}~,
\end{eqnarray}
since $\bar{D}e^a=0 $, the first term $\bar{D}K \wedge E^{D-2}$, can be reduced to a total derivative:
\begin{eqnarray}
\notag \bar{D}K\wedge E^{D-2}&=& \bar{D} \left( K\wedge E^{D-2} \right) \\ 
&=& d\left( K\wedge E^{D-2} \right)~,
\end{eqnarray}
while the second term on the left hand side of (\ref{Lagrangian1}) corresponds to the Einstein-Hilbert Lagrangian $\bar{\mathcal{L}}_{(1)}=\bar{R} \wedge E^{D-2}$. Therefore, equation (\ref{Lagrangian1}) can be rewritten as
 \begin{equation}
 \mathcal{L}_{(1)}=\bar{\mathcal{L}}_{(1)}+\langle KK \rangle \wedge E^{D-2} + d(K \wedge E^{D-2})~.
  \end{equation}
Now, we set the curvature to be null $R=0$; thus, the above expression yields
 \begin{equation}
 \bar{\mathcal{L}}_{(1)}=-\langle KK \rangle \wedge E^{D-2} - d(K \wedge E^{D-2})~.
 \end{equation}
 Writing this equation in the standard notation we have:
 \begin{eqnarray}\label{fi}
\notag  \epsilon_{a_1 a_2 \cdot\cdot\cdot  a_D}\bar{R}^{a_1 a_2}\wedge e^{a_3} \wedge \cdot\cdot\cdot \wedge e^{a_D}&=&-\epsilon_{a_1 a_2 \cdot\cdot\cdot  a_D}K^{a_1}_{\,\,\,\ c}\wedge K^{c a_2} \wedge e^{a_3} \wedge \cdot\cdot\cdot \wedge e^{a_D}\\
&&-d \left(  \epsilon_{a_1 a_2 \cdot\cdot\cdot  a_D}K^{a_1 a_2}\wedge e^{a_3} \wedge \cdot\cdot\cdot \wedge e^{a_D} \right) ~.
 \end{eqnarray}
 Because the Lagrangian $ T=-\langle KK \rangle \wedge E^{D-2}$ differs from the Einstein-Hilbert Lagrangian $\bar{\mathcal{L}}_{(1)}$ by a total derivative, both Lagrangians yield the same field equations when varying them with respect to the veilbein $e^a$ (and connection $\omega^{a}_{\,\,\ b}$, which can be set to zero for simplicity); thus, the theory defined by $T$ is called the TEGR, and it is based only on torsion. In Appendix A we show the equivalence of equations (\ref{fi}) and (\ref{dif}).
 
Notice that the equation $R^{a}_{\,\,\ b}=d\omega^{a}_{\,\,\ b}+\omega^{a}_{\,\,\ c} \wedge \omega^{c}_{\,\,\ b}=0$ can be solved trivially by performing a linear transformation of the frame and connection, and it is always possible to choose the linear transformation in such a way that the transformed local connection becomes trivial $\omega^{a}_{\,\,\ b}=0$ \cite{Obukhov:2002tm}. Also notice that $\omega^{a}_{\,\,\ b}=0$ implies that the contortion $K^{a}_{\,\,\ b}$ becomes a connection (minus the Levi-Civita connection).

\item {Teleparallel equivalent of Gauss-Bonnet gravity}

Now, we will consider the case $n=2$, i.e, the Gauss-Bonnet term. This case was studied in  \cite{Kofinas:2014owa}, where a teleparallel Lagragian equivalent to the Gauss-Bonnet Lagrangian was constructed.
In our notation, the quadratic term in the curvature of the Lanczos-Lovelock Lagrangian is written as
\begin{eqnarray} \label{Lagrangian2}
\notag \mathcal{L}_{(2)}&=& R^2 \wedge E^{D-4}\\ 
&=&(\bar{D}K+\bar{R}+\langle KK \rangle )^2\wedge E^{D-4}\\
\notag &=&\left( \left(\bar{D}K \right)^2+\bar{R}^2+\langle KK\rangle^2+2\bar{D}K \wedge \bar{R}+2\bar{D}K  \wedge\langle KK\rangle+2\bar{R}  \wedge \langle KK\rangle \right) \wedge E^{D-4}~,
\end{eqnarray}
because $\bar{D}e^a=0 $ and $\bar{D}\bar{R}^{ab}=0$, the fourth term $2\bar{D}K \wedge \bar{R} \wedge E^{D-4}$, is a total derivative:
\begin{eqnarray}
\notag 2\bar{D}K \wedge \bar{R} \wedge E^{D-4}&=& \bar{D} \left( 2 K  \wedge \bar{R} \wedge E^{D-4} \right) \\ 
&=& d\left( 2 K  \wedge \bar{R} \wedge E^{D-4} \right)~;
\end{eqnarray}
whereas the second term corresponds to the Gauss-Bonnet Lagrangian $\bar{\mathcal{L}}_{(2)}=\bar{R}^2 \wedge E^{D-4}$. Therefore, equation (\ref{Lagrangian2}) can be rewritten as
 \begin{equation} \label{GB}
 \mathcal{L}_{(2)}=\bar{\mathcal{L}}_{(2)}+ \left( \left(\bar{D}K \right)^2+\langle KK\rangle^2+2\bar{D}K  \wedge\langle KK\rangle+2\bar{R}  \wedge \langle KK\rangle \right) \wedge E^{D-4} +d\left( 2 K  \wedge \bar{R} \wedge E^{D-4} \right) ~.
  \end{equation}
 Now, replacing expressions (\ref{expressions}) in (\ref{GB}), which in our notation read
 \begin{eqnarray} \label{relacion}
  \notag \bar{D}K &=& DK-2\langle KK \rangle~, \\
  \bar{R} &=& R+\langle KK \rangle -DK~,
  \end{eqnarray}
 and setting the curvature to null $R=0$, we have that $\mathcal{L}_{(2)} = R^2 \wedge E^{D-4} = 0$, and after some straightforward calculations we obtain
 \begin{equation}
 \bar{\mathcal{L}}_{(2)}=-\left( \left( DK\right)^2 -4 DK \wedge \langle KK \rangle+3 \langle KK \rangle ^2  \right) \wedge E^{D-4}- d(2 K  \wedge \bar{R} \wedge E^{D-4})~.
 \end{equation}
 Writing this equation in the standard notation we have:
 \begin{eqnarray}
\notag  \epsilon_{a_1 a_2 \cdot\cdot\cdot  a_D}\bar{R}^{a_1 a_2} \wedge \bar{R}^{a_3 a_4} \wedge e^{a_5} \wedge \cdot\cdot\cdot \wedge e^{a_D} &=& -\epsilon_{a_1 a_2 \cdot\cdot\cdot  a_D}DK^{a_1 a_2} \wedge DK^{a_3 a_4} \wedge e^{a_5} \wedge \cdot\cdot\cdot \wedge e^{a_D}\\
 \notag &+& 4\epsilon_{a_1 a_2 \cdot\cdot\cdot  a_D}DK^{a_1 a_2} \wedge K^{a_3}_{\,\,\,\ c} \wedge K^{c a_4} \wedge e^{a_5} \wedge \cdot\cdot\cdot \wedge e^{a_D} \\
 \notag &-&3\epsilon_{a_1 a_2 \cdot\cdot\cdot  a_D}K^{a_1}_{\,\,\,\ c}  \wedge K^{c a_2} \wedge K^{a_3}_{\,\,\,\ d} \wedge K^{d a_4} \wedge e^{a_5} \wedge \cdot\cdot\cdot \wedge e^{a_D} \\
 &-& d \left( 2 \epsilon_{a_1 a_2 \cdot\cdot\cdot  a_D}K^{a_1 a_2} \wedge \bar{R}^{a_3 a_4} \wedge e^{a_5} \wedge \cdot\cdot\cdot \wedge e^{a_D} \right) ~.
 \end{eqnarray}
 The Lagrangian
 \begin{eqnarray}
\notag  T_{GB}^{(1)}&=&-\left( \left( DK\right)^2 -4 DK \wedge \langle KK \rangle+3 \langle KK \rangle ^2  \right)\wedge E^{D-4} \\
 \notag  &=&-\epsilon_{a_1 a_2 \cdot\cdot\cdot  a_D}DK^{a_1 a_2} \wedge DK^{a_3 a_4} \wedge e^{a_5} \wedge \cdot\cdot\cdot \wedge e^{a_D}\\
 \notag &+& 4\epsilon_{a_1 a_2 \cdot\cdot\cdot  a_D}DK^{a_1 a_2} \wedge K^{a_3}_{\,\,\,\ c} \wedge K^{c a_4} \wedge e^{a_5} \wedge \cdot\cdot\cdot \wedge e^{a_D} \\
  &-&3\epsilon_{a_1 a_2 \cdot\cdot\cdot  a_D}K^{a_1}_{\,\,\,\ c}  \wedge K^{c a_2} \wedge K^{a_3}_{\,\,\,\ d} \wedge K^{d a_4} \wedge e^{a_5} \wedge \cdot\cdot\cdot \wedge e^{a_D}
 \end{eqnarray}
 differs from the Gauss-Bonnet Lagrangian $\bar{\mathcal{L}}_{2}$ by a total derivative; therefore, the same field equations are obtained from both Lagrangians. Thus, the theory defined by $T_{GB}^{(1)}$ is called the teleparallel equivalent of Gauss-Bonnet gravity and it is based solely on torsion.
Additionally, we can impose the Weitzenb{\"o}ck connection by choosing $\omega^{a b}=0$ in order to simplify further the above expressions. In this case $DK^{a b}$ reduces to $dK^{a b}$. 

Also, from Equation (\ref{GB}) we see that we can integrate by parts the first term, we can integrate the third term, or both. Therefore, we can obtain other Lagrangians, besides $T_{GB}^{(1)}$, which after doing $R=0$ in the equations, they differ from the Gauss-Bonnet term by a total derivative. Now we will construct such Lagrangians explicitly. 

First, integrating by parts the first term in Equation (\ref{GB}) we obtain:
\begin{equation}
\left( \bar{D}K \right)^2 \wedge E^{D-4}=K\wedge \bar{D}^2K \wedge E^{D-4}+d(K \wedge \bar{D}K \wedge E^{D-4})~,
\end{equation}
and using $\bar{D}^2 K = 2 \langle \bar{R} K \rangle = 2 (\langle  R K  \rangle + \langle \langle  KK  \rangle  K  \rangle-\langle  DK  \rangle )$, along with the relations (\ref{relacion}) and setting $R=0$ we obtain
\begin{eqnarray}
\notag \bar{\mathcal{L}}_2 &=& -\left(  2K \wedge \langle KKK \rangle -2K \wedge \langle (DK) K \rangle -\langle KK \rangle ^2 \right) \wedge E^{D-4}\\
&& -d(2 K  \wedge \bar{R} \wedge E^{D-4} + K \wedge \bar{D}K \wedge E^{D-4})~,
\end{eqnarray}
where $\langle KKK \rangle=\langle \langle KK \rangle K \rangle$ denotes contractions of the type $K^{a}_{\,\,\,\ c} \wedge K^{c}_{\,\,\,\ d} \wedge K^{d b}$. Writing this equation in the standard notation we have:
 \begin{eqnarray}
\notag  \epsilon_{a_1 a_2 \cdot\cdot\cdot  a_D}\bar{R}^{a_1 a_2} \wedge \bar{R}^{a_3 a_4} \wedge e^{a_5} \wedge \cdot\cdot\cdot \wedge e^{a_D} &=&-2\epsilon_{a_1 a_2 \cdot\cdot\cdot  a_D}K^{a_1 a_2} \wedge K^{a_3}_{\,\,\,\ c} \wedge K^{c}_{\,\,\,\ d} \wedge K^{d a_4} \wedge e^{a_5} \wedge \cdot\cdot\cdot \wedge e^{a_D}\\
 \notag &+& 2\epsilon_{a_1 a_2 \cdot\cdot\cdot  a_D}K^{a_1 a_2} \wedge DK^{a_3}_{\,\,\,\ c} \wedge K^{c a_4} \wedge e^{a_5} \wedge \cdot\cdot\cdot \wedge e^{a_D} \\
 \notag &+&\epsilon_{a_1 a_2 \cdot\cdot\cdot  a_D}K^{a_1}_{\,\,\,\ c}  \wedge K^{c a_2} \wedge K^{a_3}_{\,\,\,\ d} \wedge K^{d a_4} \wedge e^{a_5} \wedge \cdot\cdot\cdot \wedge e^{a_D} \\
 \notag &-& d(  2 \epsilon_{a_1 a_2 \cdot\cdot\cdot  a_D}K^{a_1 a_2} \wedge \bar{R} ^{a_3 a_4} \wedge e^{a_5} \wedge \cdot\cdot\cdot \wedge e^{a_D} \\
 &+& \epsilon_{a_1 a_2 \cdot\cdot\cdot  a_D}K^{a_1 a_2} \wedge \bar{D}K^{a_3 a_4} \wedge e^{a_5} \wedge \cdot\cdot\cdot \wedge e^{a_D} )~.
 \end{eqnarray}
The Lagrangian
\begin{eqnarray}
\notag T_{GB}^{(2)} &=& -\left(  2K\wedge \langle KKK \rangle -2K \wedge \langle (DK) K \rangle -\langle KK \rangle ^2 \right) \wedge E^{D-4} \\
\notag &=& -2\epsilon_{a_1 a_2 \cdot\cdot\cdot  a_D}K^{a_1 a_2} \wedge K^{a_3}_{\,\,\,\ c} \wedge K^{c}_{\,\,\,\ d} \wedge K^{d a_4} \wedge e^{a_5} \wedge \cdot\cdot\cdot \wedge e^{a_D}\\
 \notag &+& 2\epsilon_{a_1 a_2 \cdot\cdot\cdot  a_D}K^{a_1 a_2} \wedge DK^{a_3}_{\,\,\,\ c} \wedge K^{c a_4} \wedge e^{a_5} \wedge \cdot\cdot\cdot \wedge e^{a_D} \\
 &+&\epsilon_{a_1 a_2 \cdot\cdot\cdot  a_D}K^{a_1}_{\,\,\,\ c}  \wedge K^{c a_2} \wedge K^{a_3}_{\,\,\,\ d} \wedge K^{d a_4} \wedge e^{a_5} \wedge \cdot\cdot\cdot \wedge e^{a_D}
\end{eqnarray}
also differs from $\bar{\mathcal{L}}_2$ by a total derivative; therefore, the field equations obtained by varying the resulting action are the same as Gauss-Bonnet gravity. This expression was obtained in \cite{Kofinas:2014owa}.

Second, by integrating by parts the third term in Equation (\ref{GB}) and performing the same procedure as before, the following Lagrangian is found:
\begin{eqnarray}
\notag \bar{\mathcal{L}}_2 &=& T_{GB}^{(3)} - d(2 K  \wedge \bar{R} \wedge E^{D-4}+ 2 K \wedge \langle KK \rangle  E^{D-4})~, \\
\notag T_{GB}^{(3)} &=& -\left(  (DK)^2 -6DK \wedge  \langle KK \rangle+ 7 \langle KK \rangle ^2- 4K \wedge \langle (DK) K \rangle -8K \wedge \langle KKK \rangle \right) \wedge E^{D-4} \\
\notag &=& -\epsilon_{a_1 a_2 \cdot\cdot\cdot  a_D}DK^{a_1 a_2} \wedge DK^{a_3 a_4} \wedge e^{a_5} \wedge \cdot\cdot\cdot \wedge e^{a_D}\\
 \notag &+& 6\epsilon_{a_1 a_2 \cdot\cdot\cdot  a_D}DK^{a_1 a_2} \wedge K^{a_3}_{\,\,\,\ c} \wedge K^{c a_4} \wedge e^{a_5} \wedge \cdot\cdot\cdot \wedge e^{a_D} \\
 \notag &-&7\epsilon_{a_1 a_2 \cdot\cdot\cdot  a_D}K^{a_1}_{\,\,\,\ c}  \wedge K^{c a_2} \wedge K^{a_3}_{\,\,\,\ d} \wedge K^{d a_4} \wedge e^{a_5} \wedge \cdot\cdot\cdot \wedge e^{a_D} \\
\notag &+&  4\epsilon_{a_1 a_2 \cdot\cdot\cdot  a_D}K^{a_1 a_2} \wedge DK^{a_3}_{\,\,\,\ d} \wedge K^{d a_4} \wedge e^{a_5} \wedge \cdot\cdot\cdot \wedge e^{a_D}\\
 &+& 8\epsilon_{a_1 a_2 \cdot\cdot\cdot  a_D}K^{a_1 a_2} \wedge K^{a_3}_{\,\,\,\ c} \wedge K^{c}_{\,\,\,\ d} \wedge K^{d a_4} \wedge e^{a_5} \wedge \cdot\cdot\cdot \wedge e^{a_D}~.
\end{eqnarray}
Finally, when integrating by parts both; the first and the third term in Equation (\ref{GB}), we obtain:
\begin{eqnarray}
\notag \bar{\mathcal{L}}_2 &=& T_{GB}^{(4)}  -d(2 K  \wedge \bar{R} \wedge E^{D-4} +K \wedge \bar{D}K \wedge E^{D-4} + 2 K \wedge \langle KK \rangle  E^{D-4})~, \\
\notag T_{GB}^{(4)} &=& -\left(  -6K \wedge \langle KKK \rangle +2K \wedge \langle (DK) K \rangle +3 \langle KK \rangle ^2  -2DK \wedge  \langle KK \rangle  \right) \wedge E^{D-4} \\
\notag&=& 6\epsilon_{a_1 a_2 \cdot\cdot\cdot  a_D}K^{a_1 a_2} \wedge K^{a_3}_{\,\,\,\ c} \wedge K^{c}_{\,\,\,\ d} \wedge K^{d a_4} \wedge e^{a_5} \wedge \cdot\cdot\cdot \wedge e^{a_D}\\
 \notag &-& 2\epsilon_{a_1 a_2 \cdot\cdot\cdot  a_D}K^{a_1 a_2} \wedge DK^{a_3}_{\,\,\,\ c} \wedge K^{c a_4} \wedge e^{a_5} \wedge \cdot\cdot\cdot \wedge e^{a_D} \\
 \notag &-&3\epsilon_{a_1 a_2 \cdot\cdot\cdot  a_D}K^{a_1}_{\,\,\,\ c}  \wedge K^{c a_2} \wedge K^{a_3}_{\,\,\,\ d} \wedge K^{d a_4} \wedge e^{a_5} \wedge \cdot\cdot\cdot \wedge e^{a_D} \\
 &+& 2\epsilon_{a_1 a_2 \cdot\cdot\cdot  a_D}DK^{a_1 a_2} \wedge K^{a_3}_{\,\,\,\ c} \wedge K^{c a_4} \wedge e^{a_5} \wedge \cdot\cdot\cdot \wedge e^{a_D} ~.
\end{eqnarray}

Therefore, we have found four Lagrangians $T_{GB}^{(i)}$, $i=1,2,3,4$ that differ from the Gauss-Bonnet term only by boundary terms, and thus the same field equations as Gauss-Bonnet gravity are obtained from them. Additionally, because the Gauss-Bonnet term is a topological invariant in four dimensions, so are $T_{GB}^{(i)}$, $i=1,2,3,4$. Finally, we can impose the Weitzenb{\"o}ck connection by choosing $\omega^{a b}=0$ in order to simplify further the above expressions. In this case $DK^{a b}$ reduces to $dK^{a b}$. In Appendix B we justify why it is permitted to set $\omega^{ab}=0$ in the Lagrangians in order to obtain the teleparallel equivalent of Gauss-Bonnet and Lovelock gravity in spite of the dynamics these Lagrangians contain for $\omega^{ab}$.

\item {Teleparallel equivalent of Lovelock gravity}

Here, we construct a Lagrangian which differs from the Lovelock Lagrangian solely by boundary terms. As we showed in the previous sections, different Lagrangians can be constructed differing only by boundary terms.

The term of order $n$ in the curvature of the Lanczos-Lovelock Lagrangian is written in terms of the Riemannian curvature and contortion as
\begin{eqnarray}\label{polimonio}
\notag \mathcal{L}_{(n)}&=&R^{n}\wedge E^{D-2n}\\ 
\notag &=&(\bar{D}K+\bar{R}+\langle KK \rangle )^{n}\wedge E^{D-2n}\\ 
&=& \sum_{k=0}^{n}\binom  {n} {k}(\bar{D}K)^k \wedge \sum_{j=0}^{n-k}\binom  {n-k} {j}\bar{R}^j \wedge \langle KK \rangle ^{n-k-j} \wedge E ^{D-2n}~,
\end{eqnarray}
where in the last line we have used the binomial expansion. This equation can then be expanded in the following way:
\begin{eqnarray}\label{polinomios}
\notag \mathcal{L}_{(n)}&=& \bar{R}^{n}\wedge E^{D-2n}+ \sum_{j=0}^{n-1}\binom  {n} {j}\bar{R}^j\wedge\langle KK \rangle ^{n-j} \wedge E ^{D-2n} +n \bar{D}K \wedge \sum_{j=0}^{n-2}\binom  {n-1} {j}\bar{R}^j \wedge \langle KK \rangle ^{n-1-j} \wedge E ^{D-2n} \\
&+&\sum_{k=2}^{n}\binom  {n} {k}(\bar{D}K)^k \wedge \sum_{j=0}^{n-k}\binom  {n-k} {j}\bar{R}^j \wedge \langle KK \rangle ^{n-k-j} \wedge E ^{D-2n} + n\bar{D}K \wedge \bar{R}^{n-1} \wedge E^{D-2n}~,
\end{eqnarray}
where the last term of the above equation is an exact form:
\begin{equation}
n\bar{D}K \wedge \bar{R}^{n-1} \wedge E^{D-2n}=\bar{D} \left( n K \wedge \bar{R}^{n-1}  \wedge E^{D-2n} \right)= d \left( n K \wedge \bar{R}^{n-1}  \wedge E^{D-2n} \right)~.
\end{equation}
We note that Equation (\ref{polinomios}) can be rewritten in the following form
\begin{equation}
\mathcal{L}_{(n)} = \mathcal{L}_{(n)}+\bar{\mathcal{L}}_{(n)}-\bar{R}^{n}\wedge E^{D-2n}-n\bar{D}K \wedge \bar{R}^{n-1} \wedge E^{D-2n}+\bar{D} \left( n K \wedge \bar{R}^{n-1}  \wedge E^{D-2n} \right)~,
\end{equation}
and, using Expression (\ref{relacion}) and imposing $R=0$, we get
\begin{equation}
\bar{\mathcal{L}}_{(n)}=\left( \langle  KK \rangle - DK \right)^{n-1} \wedge  \left( (1-2n) \langle KK  \rangle  -(1-n) DK \right)\wedge E^{D-2n}-d\left( n K \wedge \bar{R}^{n-1}  \wedge E^{D-2n} \right)~.
\end{equation}
Therefore, the teleparallel equivalent of Lovelock gravity is
\begin{eqnarray}
\notag T_L^{(1)} &=& \left( \langle  KK \rangle - DK \right)^{n-1} \wedge \left( (1-2n) \langle KK  \rangle  -(1-n) DK \right) \wedge E^{D-2n} \\
\notag &=& \epsilon_{a_1 a_2 \cdot\cdot\cdot  a_D}(K^{a_1 c_1}\wedge K_{c_1}^{\,\,\,a_2}-DK^{a_1 a_2})\wedge\cdot\cdot\cdot \, \wedge (K^{a_{2n-3} c_{2n-3}}\wedge K_{c_{2n-3}}^{\,\,\,\,\,\,\,\,a_{2n-2}}-DK^{a_{2n-3} a_{2n-2}})\wedge \\
&& ((1-2n)K^{a_{2n-1} c_{2n-1}}\wedge K^{\,\,\,\,\,\,\,\, a_{2n}}_{ c_{2n-1}}-(1-n)DK^{a_{2n-1} a_{2n}})\wedge e^{a_{2n+1}}\wedge\cdot\cdot\cdot\, \wedge e^{a_D}~.
\end{eqnarray}
Note that for $n=1$ and $n=2$ we recover $T$ and $T_{GB}^{(1)}$, respectively.

In order to obtain another expression for the teleparallel Lagrangian, differing of $T_L^{(1)}$ by an exact form, we will perform some integrations by parts. We begin with
\begin{eqnarray}\label{polino}
\nonumber \mathcal{L}_{(n)} &=& \sum_{j=0}^{n-1}\binom  {n} {j}\bar{R}^j\wedge \langle KK \rangle ^{n-j} \wedge E ^{D-2n}+\bar{R}^{n} \wedge E^{D-2n} \\
 &+& \sum_{k=1}^{n}\binom  {n} {k}(\bar{D}K)^k \wedge \sum_{j=0}^{n-k}\binom  {n-k} {j}\bar{R}^j \wedge \langle KK \rangle ^{n-k-j} \wedge E ^{D-2n}~,
\end{eqnarray}
then, defining
\begin{equation}
\sum_{k=1}^{n}\binom  {n} {k}\wedge (\bar{D}K)^k\wedge \sum_{j=0}^{n-k}\binom  {n-k} {j}\bar{R}^j \wedge \langle KK \rangle ^{n-k-j}\wedge E ^{D-2n}=\sum_{k=1}^{n}\binom  {n} {k}(\bar{D}K)^k \wedge S^{D-2k}~,
\end{equation}
where
\begin{equation}
S^{D-2k}=\sum_{j=0}^{n-k}\binom  {n-k} {j}\bar{R}^j\wedge \langle KK \rangle ^{n-k-j}\wedge E ^{D-2n}~,
\end{equation}
and, integrating by parts, we obtain
\begin{equation}
(\bar{D}K)^j\wedge S^{D-2j}=(j-1)K\wedge (\bar{D}K)^{j-2} \wedge \bar{D}^2K \wedge S^{D-2j}+K\wedge (\bar{D}K)^{j-1} \wedge \bar{D}(S^{D-2j})+d(K\wedge (\bar{D}K)^{j-1} \wedge S^{D-2j})~.
\end{equation}
So, by employing this, (\ref{polino}) can be written as
\begin{eqnarray}\label{polinomiow}
\nonumber \mathcal{L}_{(n)}&=& \sum_{j=0}^{n-1}\binom  {n} {j}\bar{R}^{j}\wedge\langle KK \rangle ^{n-j} \wedge E ^{D-2n}+\bar{R}^{n}\wedge E^{D-2n}\\ 
\nonumber &+&\sum_{m=1}^{n}\binom  {n} {m}K\wedge(\bar{D}K)^{m-2}\wedge \sum_{j=0}^{n-m}\binom  {n-m} {j}\bar{R}^j\wedge \{2(m-1)(\langle RK \rangle-\langle (DK)K \rangle +\langle KKK \rangle ) \wedge  \\
 \nonumber && \langle KK \rangle^{n-m-j}+ 2(n-m-j)\langle KK \rangle^{n-m-j-1}\wedge (DK-2\langle KK \rangle)\wedge (\langle (DK)K \rangle-2\langle KKK\rangle ) \} \wedge E^{D-2n}\\
&+& d\{\sum_{m=1}^{n}\binom  {n} {m}K\wedge (\bar{D}K)^{m-1}\wedge \sum_{j=0}^{n-m}\binom  {n-m} {j}\bar{R}^j\wedge \langle KK \rangle^{n-m-j}\wedge E^{D-2n}\}~,
\end{eqnarray}
where, we have used the relation
\begin{equation}
\bar{D}(\langle KK\rangle^j)=2j\langle (\bar{D}K)K\rangle \wedge \langle KK\rangle^{j-1}~.
\end{equation}
Now, in order to extract the teleparallel equivalent of Lovelock gravity $T_L$, we set the curvature to null, that is, $R=0$. In this way, Equation (\ref{polinomiow}) can be written as 
\begin{eqnarray}
\nonumber && \sum_{j=0}^{n-1}\binom  {n} {j}\bar{R}^j\wedge \langle KK \rangle ^{n-j}\wedge E ^{D-2n}+\bar{R}^n \wedge E ^{D-2n} +d(K\wedge (-\bar{D}K)^{n-1}\wedge E^{D-2n})\\
&&-2K\wedge (2\langle KK\rangle-DK)^{n-2}\wedge ((1-n)\langle (DK)K\rangle+(2n-1)\langle KKK \rangle)\wedge E^{D-2n}=0~.
\end{eqnarray}
Therefore
\begin{equation}
\bar{\mathcal{L}}_{n}=T_{L}-d(K\wedge (-\bar{D}K)^{n-1}\wedge E^{D-2n})~,
\end{equation}
where
\begin{eqnarray}
\nonumber T_L&=& - \sum_{j=0}^{n-1}\binom  {n} {j}\bar{R}^j\wedge \langle KK \rangle ^{n-j} \wedge E ^{D-2n}\\
&+& 2K\wedge ( 2\langle KK \rangle-DK)^{n-2}\wedge ((1-n)\langle (DK)K\rangle+(2n-1)\langle KKK \rangle)\wedge E^{D-2n}~.
\end{eqnarray}
Then, rewriting the sum in the following form
\begin{eqnarray}
\notag \sum_{j=0}^{n-1}\binom  {n} {j}\bar{R}^j\wedge \langle KK \rangle ^{n-j} \wedge E ^{D-2n} &=& \notag\left( \bar{R} +\langle KK \rangle \right)^n \wedge E ^{D-2n}- \bar{R}^n \wedge E ^{D-2n} \\
\notag &=& \left( 2\langle KK \rangle - DK \right)^n \wedge E ^{D-2n}\\
&-& \left(\langle KK \rangle - DK \right)^n \wedge E ^{D-2n}~,
\end{eqnarray}
we finally obtain
\begin{eqnarray}
\nonumber T_L^{(2)}&=& -\left( 2\langle KK \rangle - DK \right)^n \wedge E ^{D-2n}+ \left(\langle KK \rangle - DK \right)^n \wedge E ^{D-2n}
\\
\notag &+& 2K\wedge ( 2\langle KK \rangle-DK)^{n-2}\wedge ((1-n)\langle (DK)K\rangle+(2n-1)\langle KKK\rangle)\wedge E^{D-2n} \\
\notag &=&-\epsilon_{a_1 a_2 \cdot\cdot\cdot  a_D}(2K^{a_1 c_1}\wedge K_{c_1}^{\,\,\,a_2}-DK^{a_1 a_2})\wedge\cdot\cdot\cdot \, \wedge (2K^{a_{2n-1} c_{2n-1}}\wedge K_{c_{2n-1}}^{\,\,\,\,\,\,\,\,a_{2n}}-DK^{a_{2n-1} a_{2n}}) \wedge \\
\notag &&e^{a_{2n+1}}\wedge\cdot\cdot\cdot\, \wedge e^{a_D} +\epsilon_{a_1 a_2 \cdot\cdot\cdot  a_D}(K^{a_1 c_1}\wedge K_{c_1}^{\,\,\,a_2}-DK^{a_1 a_2})\wedge\cdot\cdot\cdot \, \wedge \\
\notag && \wedge (K^{a_{2n-1} c_{2n-1}}\wedge K_{c_{2n-1}}^{\,\,\,\,\,\,\,\,a_{2n}}-DK^{a_{2n-1} a_{2n}}) \wedge e^{a_{2n+1}}\wedge\cdot\cdot\cdot\, \wedge e^{a_D} \\
\notag &&+2 \epsilon_{a_1 a_2 \cdot\cdot\cdot  a_D}K^{a_1 a_2}\wedge(2K^{a_3 c_3}\wedge K_{c_3}^{\,\,\,a_4}-DK^{a_3 a_4})\wedge\cdot\cdot\cdot \, \wedge (2K^{a_{2n-3} c_{2n-3}}\wedge K_{c_{2n-3}}^{\,\,\,\,\,\,\,\,a_{2n-2}}-DK^{a_{2n-3} a_{2n-2}})\wedge \\
&&((1-n)DK^{a_{2n-1} c_{2n-1}}\wedge K^{\,\,\,\,\,\,\,\, a_{2n}}_{ c_{2n-1}}+(2n-1)K^{a_{2n-1} c_{2n-1}}\wedge K_{c_{2n-1}}^{\,\,\,\,\,\,\,\,c_{2n}}\wedge K_{c_{2n}}^{\,\,\,\,\,\,\,\,a_{2n}})\wedge e^{a_{2n+1}}\wedge\cdot\cdot\cdot\, \wedge e^{a_D}~.
\end{eqnarray}
Note that for $n=2$ the teleparallel equivalent of Gauss-Bonnet gravity $T_{GB}^{(4)}$ is recovered.

\end{itemize}

\section{Conclusions}

Motivated by TEGR and its recents extensions and generalizations as the so-called $f(T)$ gravity, the teleparallel equivalent of Gauss-Bonnet gravity \cite{Kofinas:2014owa, Kofinas:2014daa}, Kaluza-Klein theory for teleparallel gravity \cite{Geng:2014nfa} and scalar-torsion gravity theories \cite{Geng:2011aj, Kofinas:2015hla}, in this work we have constructed the teleparallel equivalent of Lovelock gravity $T_L$ with the vielbein and the connection, and without imposing the Weitzenb{\"o}ck connection. Then, 
we extracted the teleparallel equivalent of Lovelock gravity $T_L$ 
by setting the curvature to null as in \cite{Kofinas:2014owa}, where the teleparallel equivalent of Gauss-Bonnet gravity has been constructed. Also, we have established four possible Lagrangians for the teleparallel equivalent of Gauss-Bonnet gravity, differing among them by boundary terms. Since Lovelock Lagrangians lead to second-order equations for the metric, we expect from the relation between the metric and the vielbein (\ref{cnn}) that the teleparallel equations for the vielbein must inherit this property, containing second derivatives of the vielbein, despite the teleparallel Lagrangians also containing second derivatives of the vielbein (first derivatives of the contortion).
Nowadays, it would be interesting to study black hole solutions as well as some cosmological implications for $T_L$ and $f(T,T_L)$, to discuss whether the theory is a bad or a good candidate to describe the nature. Work in this direction is in progress.

\section*{Acknowledgments}

We would like to thank the anonymous referee for the very useful comments which help us improve the quality of our paper. This work was funded by the Comisi\'{o}n
Nacional de Ciencias y Tecnolog\'{i}a through FONDECYT Grant
11140674 (PAG). P. A. G. acknowledges the hospitality of the Universidad de La Serena where part of this work was undertaken.

\appendix

\section{}

In this appendix we show the equivalence between Equations (\ref{dif}) and (\ref{fi}). First, we expand the curvature 2-form in a coordinate basis as $\bar{R}^{a_{1}a_{2}}=\frac{1}{2} \bar{R}^{a_{1}a_{2}}_{\,\,\,\,\,\,\,\,\,\ \mu \nu} dx^{\mu} \wedge dx^{\nu}$, and furthermore, in order to switch the Latin indices to Greek indices we employ the vielbeins $\bar{R}^{a_{1}a_{2}}_{\,\,\,\,\,\,\,\,\,\ \mu \nu}=\bar{R}^{\mu_{1} \mu_{2}}_{\,\,\,\,\,\,\,\,\,\ \mu \nu} e^{a_{1}}_{\,\,\,\ \mu_{1}} e^{a_{2}}_{\,\,\,\ \mu_{2}}$, so the term in the left hand side of equation (\ref{fi}) can be written as:
\begin{eqnarray}\label{q}
\notag \epsilon_{a_1 a_2 \cdot\cdot\cdot  a_D}\bar{R}^{a_1 a_2}\wedge e^{a_3} \wedge \cdot\cdot\cdot \wedge e^{a_D}&=&\frac{1}{2}\epsilon_{a_1 a_2 \cdot\cdot\cdot  a_D}\bar{R}^{a_{1}a_{2}}_{\,\,\,\,\,\,\,\,\,\ \mu \nu} dx^{\mu} \wedge dx^{\nu}\wedge e^{a_3} \wedge \cdot\cdot\cdot \wedge e^{a_D} \\
\notag &=& \frac{1}{2}\epsilon_{a_1 a_2 \cdot\cdot\cdot  a_D}\bar{R}^{\mu_{1}\mu_{2}}_{\,\,\,\,\,\,\,\,\,\ \mu \nu} e^{a_{1}}_{\,\,\,\ \mu_{1}} e^{a_{2}}_{\,\,\,\ \mu_{2}} dx^{\mu} \wedge dx^{\nu}\wedge e^{a_3} \wedge \cdot\cdot\cdot \wedge e^{a_D} \\
\notag &=& \frac{1}{2}\epsilon_{a_1 a_2 \cdot\cdot\cdot  a_D}\bar{R}^{\mu_{1}\mu_{2}}_{\,\,\,\,\,\,\,\,\,\ \mu \nu} e^{a_{1}}_{\,\,\,\ \mu_{1}} e^{a_{2}}_{\,\,\,\ \mu_{2}} e^{a_3}_{\,\,\,\ \mu_{3}} \cdot\cdot\cdot e^{a_D}_{\,\,\,\ \mu_{D}} dx^{\mu} \wedge dx^{\nu}\wedge dx^{\mu_{3}}  \wedge \cdot\cdot\cdot \wedge dx^{\mu_{D}} \\
\notag &=& \frac{1}{2} \sqrt{-g} \bar{R}^{\mu_{1}\mu_{2}}_{\,\,\,\,\,\,\,\,\,\ \mu \nu} \epsilon_{\mu_{1} \mu_{2} \cdot \cdot \cdot \mu_{D}} dx^{\mu} \wedge dx^{\nu}\wedge dx^{\mu_{3}}  \wedge \cdot\cdot\cdot \wedge dx^{\mu_{D}} \\
\notag &=& \frac{1}{2} \sqrt{-g} \bar{R}^{\mu_{1}\mu_{2}}_{\,\,\,\,\,\,\,\,\,\ \mu \nu} \epsilon_{\mu_{1} \mu_{2} \cdot \cdot \cdot \mu_{D}} \epsilon^{\mu \nu \mu_{3} \cdot \cdot \cdot \mu_{D}}dx^{0} \wedge dx^{1} \wedge \cdot\cdot\cdot \wedge dx^{D-1} \\
\notag &=& \frac{1}{2} \sqrt{-g} \bar{R}^{\mu_{1}\mu_{2}}_{\,\,\,\,\,\,\,\,\,\ \mu \nu} \delta^{\mu \nu \mu_{3} \cdot \cdot \cdot \mu_{D}}_{\mu_{1} \mu_{2} \cdot \cdot \cdot \mu_{D}}d^{D}x \\
\notag &=& \frac{1}{2} (D-2)! \sqrt{-g} \bar{R}^{\mu_{1}\mu_{2}}_{\,\,\,\,\,\,\,\,\,\ \mu \nu} \delta^{\mu \nu}_{\mu_{1} \mu_{2}}d^{D}x \\
&=& (D-2)! \sqrt{-g} \bar{R} d^{D}x~.
\end{eqnarray}
In going from the second to third line we use of the expansion $e^{a}=e^{a}_{\,\ \mu} dx^{\mu}$; in going from the third to fourth line we use of $\epsilon_{a_1 a_2 \cdot\cdot\cdot  a_D}e^{a_{1}}_{\,\,\,\ \mu_{1}} e^{a_{2}}_{\,\,\,\ \mu_{2}} e^{a_3}_{\,\,\,\ \mu_{3}} \cdot\cdot\cdot e^{a_D}_{\,\,\,\ \mu_{D}} = \sqrt{-g} \epsilon_{\mu_{1} \mu_{2} \cdot \cdot \cdot \mu_{D}}$; in going from the fourth to fifth line we use $dx^{\mu} \wedge dx^{\nu}\wedge dx^{\mu_{3}}  \wedge \cdot\cdot\cdot \wedge dx^{\mu_{D}}=\epsilon^{\mu \nu \mu_{3} \cdot \cdot \cdot \mu_{D}}dx^{0} \wedge dx^{1} \wedge \cdot\cdot\cdot \wedge dx^{D-1}=\epsilon^{\mu \nu \mu_{3} \cdot \cdot \cdot \mu_{D}}d^{D}x$, and finally we use $\epsilon_{\mu_{1} \mu_{2} \cdot \cdot \cdot \mu_{D}} \epsilon^{\mu \nu \mu_{3} \cdot \cdot \cdot \mu_{D}}=\delta^{\mu \nu \mu_{3} \cdot \cdot \cdot \mu_{D}}_{\mu_{1} \mu_{2} \cdot \cdot \cdot \mu_{D}}=(D-2)! \delta^{\mu \nu}_{\mu_{1} \mu_{2}}=(D-2)! (\delta^{\mu}_{\mu_{1}}\delta^{\nu}_{\mu_{2}}-\delta^{\mu}_{\mu_{2}}\delta^{\nu}_{\mu_{1}})$.

In a similar way we expand the first term on the right-hand side of Equation (\ref{fi}). First, we express the contortion in terms of the torsion, which are related by $T^{a}=K^{a}_{\,\ b} \wedge e^{b}$, so:
\begin{equation}\label{product}
K^{ab}=\frac{1}{2} \left(i^{a}(T^{b})-i^{b}(T^{a})+i^{b}i^{a}(T^{c})\wedge e_{c} \right)~,
\end{equation}
where $i^{a} \equiv i^{e^{a}}$ is the interior product with respect to the vielbein.

Now, expanding the torsion in the orthonormal base $T^{a}=\frac{1}{2}T^{a}_{\,\ f g} e^{f} \wedge e^{g}$ and using (\ref{product}) we obtain
\begin{eqnarray}
\notag K^{ab}&=&\frac{1}{2} ( -T^{ab}_{\,\,\,\ c} +T^{ba}_{\,\,\,\ c} +T_{c}^{\,\ ab} ) e^{c} \\
&=& \frac{1}{2} ( -T^{ab}_{\,\,\,\ \mu} +T^{ba}_{\,\,\,\ \mu} +T_{\mu}^{\,\ ab} ) dx^{\mu}~.
\end{eqnarray}
Thus, using this expression we find:
\begin{eqnarray} \label{redu}
\notag \epsilon_{a_1 a_2 \cdot\cdot\cdot  a_D}K^{a_1}_{\,\,\,\ c}\wedge K^{c a_2} \wedge e^{a_3} \wedge \cdot\cdot\cdot \wedge e^{a_D} &=&\frac{1}{4} \epsilon_{a_1 a_2 \cdot\cdot\cdot  a_D} ( -T^{a_{1}}_{\,\,\,\ c \mu_{1}} +T_{c \,\,\,\, \mu_{1}}^{\,\, a_{1}}+T_{\mu_{1}\,\,\, c}^{\,\ a_{1}}) ( -T^{c a_{2}}_{\,\,\,\,\,\,\,\,\, \mu_{2}} +T^{a_{2} c}_{\,\,\,\,\,\,\,\,\, \mu_{2}} +T_{\mu_{2}}^{\,\ c a_{2}} ) dx^{\mu_{1}} \wedge dx^{\mu_{2}} \\
\notag && \wedge  e^{a_3} \wedge \cdot\cdot\cdot \wedge e^{a_D} \\
\notag &=& \frac{1}{4} \epsilon_{a_1 a_2 \cdot\cdot\cdot  a_D} ( -T^{\nu_{1}}_{\,\,\,\ c \mu_{1}} +T_{c \,\,\,\, \mu_{1}}^{\,\, \nu_{1}}+T_{\mu_{1}\,\,\, c}^{\,\ \nu_{1}}) ( -T^{c \nu_{2}}_{\,\,\,\,\,\,\,\,\, \mu_{2}} +T^{\nu_{2} c}_{\,\,\,\,\,\,\,\,\, \mu_{2}} +T_{\mu_{2}}^{\,\ c \nu_{2}} ) e^{a_{1}}_{\,\,\,\, \nu_{1}} e^{a_{2}}_{\,\,\,\, \nu_{2}} \\
&& dx^{\mu_{1}} \wedge dx^{\mu_{2}} \wedge  e^{a_3} \wedge \cdot\cdot\cdot \wedge e^{a_D}~.
\end{eqnarray}

In order to simplify the writing, we define
\begin{eqnarray}
\notag T^{\nu_{1} \nu_{2}}_{\mu_{1} \mu_{2}}&=&( -T^{\nu_{1}}_{\,\,\,\ c \mu_{1}} +T_{c \,\,\,\, \mu_{1}}^{\,\, \nu_{1}}+T_{\mu_{1}\,\,\, c}^{\,\ \nu_{1}}) ( -T^{c \nu_{2}}_{\,\,\,\,\,\,\,\,\, \mu_{2}} +T^{\nu_{2} c}_{\,\,\,\,\,\,\,\,\, \mu_{2}} +T_{\mu_{2}}^{\,\ c \nu_{2}} ) \\
\notag &=& T^{\nu_{1}}_{\,\,\,\ c \mu_{1}}T^{c \nu_{2}}_{\,\,\,\,\,\,\,\,\, \mu_{2}} -T^{\nu_{1}}_{\,\,\,\ c \mu_{1}}T^{\nu_{2} c}_{\,\,\,\,\,\,\,\,\, \mu_{2}} -T^{\nu_{1}}_{\,\,\,\ c \mu_{1}}T_{\mu_{2}}^{\,\ c \nu_{2}}-T_{c \,\,\,\, \mu_{1}}^{\,\, \nu_{1}}T^{c \nu_{2}}_{\,\,\,\,\,\,\,\,\, \mu_{2}} +T_{c \,\,\,\, \mu_{1}}^{\,\, \nu_{1}}T^{\nu_{2} c}_{\,\,\,\,\,\,\,\,\, \mu_{2}} +T_{c \,\,\,\, \mu_{1}}^{\,\, \nu_{1}}T_{\mu_{2}}^{\,\ c \nu_{2}} \\
&&-T_{\mu_{1}\,\,\, c}^{\,\ \nu_{1}}T^{c \nu_{2}}_{\,\,\,\,\,\,\,\,\, \mu_{2}} +T_{\mu_{1}\,\,\, c}^{\,\ \nu_{1}}T^{\nu_{2} c}_{\,\,\,\,\,\,\,\,\, \mu_{2}} +T_{\mu_{1}\,\,\, c}^{\,\ \nu_{1}}T_{\mu_{2}}^{\,\ c \nu_{2}}~,
\end{eqnarray}
so, Equation (\ref{redu}) becomes
\begin{eqnarray}\label{qq}
\notag \epsilon_{a_1 a_2 \cdot\cdot\cdot  a_D}K^{a_1}_{\,\,\,\ c}\wedge K^{c a_2} \wedge e^{a_3} \wedge \cdot\cdot\cdot \wedge e^{a_D} &=& \frac{1}{4} \epsilon_{a_1 a_2 \cdot\cdot\cdot  a_D} T^{\nu_{1} \nu_{2}}_{\mu_{1} \mu_{2}} e^{a_{1}}_{\,\,\,\, \nu_{1}} e^{a_{2}}_{\,\,\,\, \nu_{2}} dx^{\mu_{1}} \wedge dx^{\mu_{2}} \wedge  e^{a_3} \wedge \cdot\cdot\cdot \wedge e^{a_D} \\
\notag &=& \frac{1}{4} \epsilon_{a_1 a_2 \cdot\cdot\cdot  a_D} T^{\nu_{1} \nu_{2}}_{\mu_{1} \mu_{2}} e^{a_{1}}_{\,\,\,\, \nu_{1}} e^{a_{2}}_{\,\,\,\, \nu_{2}} e^{a_{3}}_{\,\,\,\, \mu_{3}} \cdot \cdot \cdot e^{a_{D}}_{\,\,\,\, \mu_{D}} dx^{\mu_{1}} \wedge dx^{\mu_{2}} \wedge \cdot\cdot\cdot \wedge dx^{\mu_{D}} \\
\notag &=& \frac{1}{4} \sqrt{-g} T^{\nu_{1} \nu_{2}}_{\mu_{1} \mu_{2}} \epsilon_{\nu_{1} \nu_{2} \mu_{3} \cdot \cdot \cdot \mu_{D}} dx^{\mu_{1}} \wedge dx^{\mu_{2}} \wedge \cdot\cdot\cdot \wedge dx^{\mu_{D}} \\
\notag &=& \frac{1}{4} \sqrt{-g} T^{\nu_{1} \nu_{2}}_{\mu_{1} \mu_{2}} \epsilon_{\nu_{1} \nu_{2} \mu_{3} \cdot \cdot \cdot \mu_{D}} \epsilon^{\mu_{1} \mu_{2} \cdot \cdot \cdot \mu_{D}}dx^{0} \wedge dx^{1} \wedge \cdot\cdot\cdot \wedge dx^{D-1} \\
\notag &=& \frac{1}{4} \sqrt{-g} T^{\nu_{1} \nu_{2}}_{\mu_{1} \mu_{2}} \delta^{\mu_{1} \mu_{2} \cdot \cdot \cdot \mu_{D}}_{\nu_{1} \nu_{2} \mu_{3} \cdot \cdot \cdot \mu_{D}}dx^{0} \wedge dx^{1} \wedge \cdot\cdot\cdot \wedge dx^{D-1} \\
\notag &=& \frac{1}{4} (D-2)! \sqrt{-g} T^{\nu_{1} \nu_{2}}_{\mu_{1} \mu_{2}} \delta^{\mu_{1} \mu_{2}}_{\nu_{1} \nu_{2}}d^{D}x \\
\notag &=& \frac{1}{4} (D-2)! \sqrt{-g} (T^{\mu_{1} \mu_{2}}_{\mu_{1} \mu_{2}}-T^{\mu_{1} \mu_{2}}_{\mu_{2} \mu_{1}}) d^{D}x \\
\notag &=& (D-2)! \sqrt{-g} (-T_{\mu_{1} c}^{\,\,\,\,\,\,\,\, \mu_{1}}T^{\mu_{2} c}_{\,\,\,\,\,\,\,\,\,\, \mu_{2}}+\frac{1}{2}T^{\mu_{1} c \mu_{2}}T_{\mu_{1} c \mu_{2}}+\frac{1}{4} T^{\mu_{1} \mu_{2} c}T_{\mu_{1} \mu_{2}  c}) d^{D}x \\
\notag &=& (D-2)! \sqrt{-g} (-T_{\alpha \gamma}^{\,\,\,\,\, \alpha}T^{\beta \gamma}_{\,\,\,\,\,\ \beta}+\frac{1}{2}T^{\alpha \gamma \beta}T_{\alpha \gamma \beta}+\frac{1}{4} T^{\alpha \beta \gamma}T_{\alpha \beta  \gamma}) d^{D}x \\
&=& (D-2)! \sqrt{-g} T d^{D}x~,
\end{eqnarray}
where the torsion scalar is defined as $T=-T_{\alpha \gamma}^{\,\,\,\,\, \alpha}T^{\beta \gamma}_{\,\,\,\,\,\ \beta}+\frac{1}{2}T^{\alpha \gamma \beta}T_{\alpha \gamma \beta}+\frac{1}{4} T^{\alpha \beta \gamma}T_{\alpha \beta  \gamma}$.

The boundary term yields
\begin{eqnarray}\label{qqq}
\notag d(\epsilon_{a_{1} \cdot \cdot \cdot a_{D}} K^{a_{1} a_{2}} \wedge e^{a_{3}} \wedge \cdot \cdot \cdot \wedge e^{a_{D}})&=& \frac{1}{2} d(\epsilon_{a_{1} \cdot \cdot \cdot a_{D}} (-T^{\mu_{1} \mu_{2}}_{\,\,\,\,\,\,\,\,\,\,\,\,\, \nu}+T^{\mu_{2} \mu_{1}}_{\,\,\,\,\,\,\,\,\,\,\,\,\, \nu}+T_{\nu}^{\,\, \mu_{1} \mu_{2}}) e^{a_{1}}_{\,\,\,\, \mu_{1}} e^{a_{2}}_{\,\,\,\, \mu_{2}} dx^{\nu} \wedge e^{a_{3}} \wedge \cdot \cdot \cdot \wedge e^{a_{D}}) \\
\notag &=& \frac{1}{2} d(\epsilon_{a_{1} \cdot \cdot \cdot a_{D}} (-T^{\mu_{1} \mu_{2}}_{\,\,\,\,\,\,\,\,\,\,\,\,\, \nu}+T^{\mu_{2} \mu_{1}}_{\,\,\,\,\,\,\,\,\,\,\,\,\, \nu}+T_{\nu}^{\,\, \mu_{1} \mu_{2}}) e^{a_{1}}_{\,\,\,\, \mu_{1}} \cdot \cdot \cdot e^{a_{D}}_{\,\,\,\, \mu_{D}} dx^{\nu}\wedge dx^{\mu_{3}} \wedge \cdot \cdot \cdot \wedge dx^{\mu_{D}}) \\
\notag &=& \frac{1}{2} d(\epsilon_{a_{1} \cdot \cdot \cdot a_{D}} (-T^{\mu_{1} \mu_{2}}_{\,\,\,\,\,\,\,\,\,\,\,\,\, \nu}+T^{\mu_{2} \mu_{1}}_{\,\,\,\,\,\,\,\,\,\,\,\,\, \nu}+T_{\nu}^{\,\, \mu_{1} \mu_{2}}) e^{a_{1}}_{\,\,\,\, \mu_{1}} \cdot \cdot \cdot e^{a_{D}}_{\,\,\,\, \mu_{D}} dx^{\nu} \wedge dx^{\mu_{3}}\wedge \cdot \cdot \cdot \wedge dx^{\mu_{D}}) \\
\notag &=& \frac{1}{2} d( \sqrt{-g}(-T^{\mu_{1} \mu_{2}}_{\,\,\,\,\,\,\,\,\,\,\,\,\, \nu}+T^{\mu_{2} \mu_{1}}_{\,\,\,\,\,\,\,\,\,\,\,\,\, \nu}+T_{\nu}^{\,\, \mu_{1} \mu_{2}}) \epsilon_{\mu_{1} \cdot \cdot \cdot \mu_{D}} dx^{\nu} \wedge dx^{\mu_{3}}\wedge \cdot \cdot \cdot \wedge dx^{\mu_{D}}) \\
\notag &=& \frac{1}{2} \partial_{\mu}( \sqrt{-g}(-T^{\mu_{1} \mu_{2}}_{\,\,\,\,\,\,\,\,\,\,\,\,\, \nu}+T^{\mu_{2} \mu_{1}}_{\,\,\,\,\,\,\,\,\,\,\,\,\, \nu}+T_{\nu}^{\,\, \mu_{1} \mu_{2}})) \epsilon_{\mu_{1} \cdot \cdot \cdot \mu_{D}} dx^{\mu} \wedge dx^{\nu} \wedge dx^{\mu_{3}}\wedge \cdot \cdot \cdot \wedge dx^{\mu_{D}} \\
\notag &=& \frac{1}{2} \partial_{\mu}( \sqrt{-g}(-T^{\mu_{1} \mu_{2}}_{\,\,\,\,\,\,\,\,\,\,\,\,\, \nu}+T^{\mu_{2} \mu_{1}}_{\,\,\,\,\,\,\,\,\,\,\,\,\, \nu}+T_{\nu}^{\,\, \mu_{1} \mu_{2}})) \epsilon_{\mu_{1} \cdot \cdot \cdot \mu_{D}} \epsilon^{\mu \nu \mu_{3} \cdot \cdot \cdot \mu_{D}} dx^{0} \wedge dx^{1} \wedge \cdot \cdot \cdot \wedge dx^{D-1} \\
\notag &=& \frac{1}{2} (D-2)! \partial_{\mu}( \sqrt{-g}(-T^{\mu_{1} \mu_{2}}_{\,\,\,\,\,\,\,\,\,\,\,\,\, \nu}+T^{\mu_{2} \mu_{1}}_{\,\,\,\,\,\,\,\,\,\,\,\,\, \nu}+T_{\nu}^{\,\, \mu_{1} \mu_{2}})) \delta{\mu \nu}_{\mu_{1} \mu_{2}} d^{D}x \\
&=& (D-2)! \partial_{\mu}( 2 \sqrt{-g}T^{\nu \mu}_{\,\,\,\,\,\,\,\, \nu}) d^{D}x~.
\end{eqnarray}

Notice that the factor $(D-2)!$ appearing in Equations (\ref{q}), (\ref{qq}) and (\ref{qqq}) is canceled out with the one contained in $\kappa$. The other Lovelock terms can be expanded in a similar way.

\section{}

A formal way to proceed in obtaining the teleparallel equivalent of Lovelock gravity is to include the constraint of zero curvature $R^{ab}=0$ as a Lagrange multiplier in the gravitational action. We follow the arguments of reference \cite{Blagojevic:2002du}; there, the author has considered the case of TEGR.

As we have shown in section IV, using the splitting $\omega^{ab}=\bar{\omega}^{ab}+K^{ab}$, the Lagrangian $L_{(q)}(e ,\omega)$ defined in Equation (\ref{polinomio}), which depends on the vielbein $e^{a}$ and the general connection $\omega^{ab}$, can be decomposed in three terms: a term $\bar{L}_{(q)}(e)$ that depends only on the Levi-Civita connection $\bar{\omega}^{ab}$ and the vielbein (the Lovelock Lagrangian), a term $-T_{(q)}(e ,\omega)$ that depends on the vielbein and the connection $\omega^{ab}$, and a boundary term $B_{(q)}$:
\begin{equation}\label{splitting}
L_{(q)}(e ,\omega)=\bar{L}_{(q)}(e)-T_{(q)}(e ,\omega)+dB_{(q)}~.
\end{equation}
Note that by making integrations by parts, different $T_{(q)}$ can be obtained. 
Now, we will show that by imposing the curvature to vanish $R^{ab}=0$, the Lagrangian $T_{(q)}(e ,\omega)$ is the teleparallel equivalent of the Lovelock Lagrangian $\bar{L}_{(q)}(e)$. In order to do this, we consider the following action with the constraint $R^{ab}=0$ in the action through a Lagrange multiplier, i.e.,
\begin{equation}
S= \int \kappa c_{q}^{k}T_{(q)}(e ,\omega)+\lambda_{ab}\wedge R^{ab}~,
\end{equation}
where $\lambda_{ab}$ is a $(D-2)$-form field antisymmetric in indices $a$ and $b$.
Varying the above action with respect to the vielbein, spin connection and Lagrange multiplier, the following field equations are obtained, respectively:
\begin{equation}\label{B1}
\frac{\delta T_{(q)}(e ,\omega)}{\delta e^{a}} = 0~,
\end{equation}
\begin{equation}\label{B2}
\frac{\delta T_{(q)}(e ,\omega)}{\delta \omega^{ab}}+D \lambda_{ab} = 0~,
\end{equation}
\begin{equation}\label{B3}
R^{ab} = 0~,
\end{equation}
where we have made $\kappa c_{q}^{k}=1$ in order to simplify the writing of equations. The first equation is a dynamical equation for the vielbein; the second equation only determines the Lagrange multipliers $\lambda_{ab}$ as we will discuss below, and the last equation defines the teleparallel geometry. Now, from (\ref{splitting}) we can express the variations of $T_{(q)}(e ,\omega)$ in terms of variations of $L_{(q)}(e ,\omega)$ and $\bar{L}_{(q)}(e)$ as follows
\begin{equation}
\frac{\delta T_{(q)}(e ,\omega)}{\delta e^{a}} = \frac{\delta \bar{L}_{(q)}(e)}{\delta e^{a}}-\frac{\delta L_{(q)}(e, \omega)}{\delta e^{a}}~,
\end{equation}
\begin{equation}
\frac{\delta T_{(q)}(e ,\omega)}{\delta \omega^{ab}} = -\frac{\delta L_{(q)}(e ,\omega)}{\delta \omega^{ab}}~.
\end{equation}
So, using these expressions, the equations of motion (\ref{B1}), (\ref{B2}) and (\ref{B3}) can be written as
\begin{equation}\label{primera}
\frac{\delta T_{(q)}(e ,\omega)}{\delta e^{a}} =  \frac{\delta \bar{L}_{(q)}(e)}{\delta e^{a}}-\frac{\delta L_{(q)}(e, \omega)}{\delta e^{a}} = 0 ~,
\end{equation}
\begin{equation}\label{segunda}
-\frac{\delta L_{(q)}(e ,\omega)}{\delta \omega^{ab}}+D \lambda_{ab} = 0~,
\end{equation}
\begin{equation}\label{third}
R^{ab} = 0~.
\end{equation}
Moreover, the variations of $L_{(q)}(e, \omega)$ with respect to $e^{a}$ and $\omega^{ab}$ are given respectively by:
\begin{equation}\label{num1}
\frac{\delta L_{(q)}(e ,\omega)}{\delta e^{a}} =  (D-2q) \epsilon_{a a_{1} \cdot \cdot \cdot a_{D-1}}R^{a_{1} a_{2}} \wedge \cdot \cdot \cdot \wedge R^{a_{2q-1} a_{2q}} \wedge e^{a_{2q+1}} \wedge \cdot \cdot \cdot \wedge e^{a_{D-1}}  ~,
\end{equation}
\begin{equation}\label{num2}
\frac{\delta L_{(q)}(e ,\omega)}{\delta \omega^{ab}} =  -q(D-2q) \epsilon_{ab a_{3} \cdot \cdot \cdot a_{D}}R^{a_{3} a_{4}} \wedge \cdot \cdot \cdot \wedge R^{a_{2q-1} a_{2q}} \wedge T^{a_{2q+1}} \wedge e^{a_{2q+2}} \wedge \cdot \cdot \cdot \wedge e^{a_{D}}  ~.
\end{equation}
By taking the covariant derivative to the second field equation (\ref{segunda}) and employing (\ref{num2}), together with the Bianchi identities and the relation $D^2 \lambda_{ab}=R_{ac} \wedge \lambda^{c}_{\,\,\, b}-\lambda_{ac} \wedge R^{c}_{\,\,\, b}$, we obtain the following consistency condition
\begin{eqnarray}
\notag && q(D-2q) \epsilon_{ab a_{3} \cdot \cdot \cdot a_{D}}R^{a_{3} a_{4}} \wedge \cdot \cdot \cdot \wedge R^{a_{2q-1} a_{2q}} \wedge R^{a_{2q+1}}_{\,\,\,\,\,\,\,\,\,\,\,\,\,\ c} \wedge e^{c} \wedge e^{a_{2q+2}} \wedge \cdot \cdot \cdot \wedge e^{a_{D}} \\
\notag && +q(D-2q)(D-2q-1) \epsilon_{ab a_{3} \cdot \cdot \cdot a_{D}}R^{a_{3} a_{4}} \wedge \cdot \cdot \cdot \wedge R^{a_{2q-1} a_{2q}} \wedge T^{a_{2q+1}} \wedge T^{a_{2q+2}} \wedge e^{2q+3} \wedge \cdot \cdot \cdot \wedge e^{a_{D}} \\
&& +R_{ac} \wedge \lambda^{c}_{\,\,\, b}-\lambda_{ac} \wedge R^{c}_{\,\,\, b} =0 ~,
\end{eqnarray}
which is satisfied due to the third field equation $R^{ab}=0$. In a similar way, it can be shown that higher derivatives of (\ref{segunda}) are also satisfied. Thus, the only role of equation ($\ref{segunda}$) is to determine the Lagrange multipliers. Therefore, the non-trivial dynamic is completely contained in the first field equation (\ref{primera}), which, after using (\ref{third}) and (\ref{num1}) reduces to:
\begin{equation}
\frac{\delta T_{(q)}(e ,\omega)}{\delta e^{a}} =  \frac{\delta \bar{L}_{(q)}(e)}{\delta e^{a}} = 0 ~.
\end{equation}
This expression shows that the same equations of motion of Lovelock gravity are obtained from Lagrangian $T_{(q)}(e ,\omega)$, once the teleparallel condition is imposed. Furthermore, the teleparallel condition $R^{ab} \equiv d\omega^{ab}+\omega^{a}_{\,\,\,c} \wedge \omega^{cb}=0$ allows us to choose the gauge $\omega^{ab}=0$, i.e., the Weitzenb{\"o}ck connection. Moreover, the above analysis shows that the teleparallel theory may also be described by imposing the gauge condition $\omega^{ab}=0$ directly on the action \cite{Blagojevic:2002du}.

\vskip 7cm

\end{document}